# Continuous Frequency Controllable Nano-electromechanical Systems Based on Multiwalled Carbon Nanotubes

Quan-shui Zheng[1,2*], Zhiping Xu[1], Adrian Neild[2] and Tuck Wah Ng[2]

[1]*Department of Engineering Mechanics, Tsinghua University, Beijing, China*
[2]*Department of Mechanical Engineering, Monash University, Melbourne, Australia*

*We demonstrate a class of model nano-electromechanical systems (NEMS) based on multiwalled carbon nanotubes (MWNTs) which has longer inner cores coaxially oscillating inside their respective shorter outer shell holders and can operate at continuously controllable frequencies up to the gigahertz range when fuelled by AC electric fields. Its additional attributes include much larger oscillation amplitudes and forces and much lower rates of thermal dissipation (Q-factor = $10^5$) and air damping (Q-factor = $10^4$~$10^5$) than those of nano-beam based NEMS. A crucial feature of the conceived model NEMS is that after having tuned the electric field frequency to any prescribed value within a permitted range, the NEMS will respond quickly (in sub-nanoseconds) at the same oscillation frequency. These merits, when contrasted with the nano-beam resonators developed so far, make it a better potential candidate for the ongoing miniaturization progress from micro- to nano-electromechanical systems.*

---

[*] To whom correspondence should be addressed. E-mails: zhengqs@tsinghua.edu.cn

Nanotechnology is rapidly progressing from the passive and active nanostructures to nano- and molecular systems, with an estimated market value toward US$ 1,000 billion by the year 2015[1]. One of the most exciting areas in this field is nano-electromechanical systems (NEMS) which holds the promise of devices providing atomic scale position precision and ultra-high mechanical frequencies in the gigahertz (GHz) range. This is regarded as the holy grail in miniaturization of electromechanical devices[2]. Carbon nanotubes are quasi-one dimensional nanostructures with many extreme and fascinating mechanical and electronic properties. For instance, Yu *et al.*[3] observed extra-low intershell shear strengths in the order of 0.08 ~ 0.3 MPa for MWNTs. This property imbues MWNTs with great mobility for intershell sliding motion. Cumings and Zettl[4] demonstrated controlled and self-reversible intershell telescopic extension. Based on these experimental observations, Zheng and Jiang[5] proposed a GHz mechanical linear oscillator based on MWNTs, which was the first of many such ideas that came along[6]. The operation of such an oscillator is simple - the inner core is extruded from one end of a two ends opened MWNT via external control, retracts rapidly into the outer shell holder upon release, decelerates and rebounds by intershell van der Waals interaction from the opposite end, and then repeats the motion such that it linearly or axially oscillates about the outer shell holder. Over the pass five years there has been intensive theoretical and experimental study on GHz oscillators and related topics. The first nano-device with measured GHz mechanical frequency was reported by Huang *et al.*[7]. It was a SiC nanocrystalline beam resonated laterally due to elasticity (hence dubbed *elas-GHz resonators*) and characterized by an electric microwave-network analyser. However, linear GHz oscillators due to the van der Waals interaction, called *vdw-GHz oscillators*, have not been realized yet, despite many experimental attempts[4, 8-10]. These attempts give rise to the following questions: (1) What are the major obstacles that prevent the realization of vdw-GHz oscillators? (2) Would vdw-GHz oscillators have substantially important advantages over elas-GHz resonators? (3) If yes, how to

conceive a new type of vdw-GHz oscillator for easier and faster realization? We will attempt to answer all of these questions here.

Although the ability to control and detect mechanical GHz frequencies of nanodevices is the major technical challenge[9,10], a more fundamental obstacle is the rapidly increased thermal or phonon damping with size shrinkage[11]. There have been copious studies[12-15], using molecular dynamics simulations to understand the mechanism of energy dissipation in double-walled carbon nanotube (DWNT)-based oscillators with lengths of several to tens of nanometers and diameters of about 1 nm. Figure 1(a) illustrates a typical DWNT oscillator investigated most in previous MD simulations that has the same inner and outer shell lengths $L$ and, consequently, a V-shaped intershell van der Waals potential (Figure 1(b)). Furthermore, the oscillation amplitude is equal to the cross-end distance $\Delta$. The MD simulation results revealed that the oscillating amplitudes always diminished within a few to dozens of nanoseconds. This observation may have the implication that even though axial oscillations might occur in the previous experiments, their oscillation life spans would be too short to be detectable. This may explain why vdw-GHz oscillators have yet to be experimentally verified. Compared with the influences of various factors, such as length[13,14,16], temperature and chirality[12], defect[17,18], cross-end distance[15] and trans-phonon effects[19,20], we conclude that the last two factors would be the main mechanisms of exciting non-translational motion modes and thus leading to high phonon damping.

To arrive at a conclusive comparison on the merits of vdw-GHz oscillators and elas-GHz resonators, we need to first give a reasonable list of performance features. For actuators, the most important features are their work and power densities, frequencies and quality or $Q$-factors[11]. We compare these capabilities in relation to force, displacement, thermal and air $Q$-factors as the two nano-devices have similar sizes and frequencies. Figure 1 show three possible configurations of DWNT oscillators and their corresponding intershell van der Waals potentials as functions of axial moving distances $x$ of the mass centers of the inner shells versus those of the outer shells. If the outer shells are fixed, then the oscillating frequencies[5,21] can be reformatted into the following unified expressions

$$f_{vdw}^{II} = \frac{V_{vdw}}{a+2\Delta}\sqrt{\frac{\Delta}{8L}} = \frac{v_{max}}{4(a+2\Delta)} \tag{1}$$

for the two sub-types of the so-called Type II oscillators that have different inner and outer shell lengths as illustrated in Figures 1(c)-(f), where $L$ denotes the inner shell length, $a$ is the half of the shell length differences, $\Delta$ is the cross-end distance, $v_{max} = V_{vdw}(2\Delta/L)^{1/2}$ is the inner core maximum traveling speed, and $V_{vdw} = (\Pi/m_0)^{1/2}$ is approximately equal to 660 m/s for typical DWNT oscillators and decreases with increase in the inner core diameter and wall number. We define $V_{vdw}$ as the van der Waals speed because it is determined by the van der Waals energy $\Pi$ of a single carbon atom on the inner core surface interacting with all (infinitively long) outer shell atoms and the carbon atom mass $m_0$. For Type-I oscillators that have equal inner and outer shell lengths (Figures 1(a)-(b)), the frequency expression is reduced to $f_{vdw}^{I} = V_{vdw}/(32L\Delta)^{1/2}$. In comparison, the resonant frequencies of, for example, a solid cylindrical cantilever nanobeam (Figure 2(b)) can be expressed in the form $f_{elas} = (\beta^2 V_{elas}/8\pi)(D/L^2)$, where $\beta$ denotes the resonant mode parameter with the leading values $\beta_1 = 1.875$, $\beta_2 = 4.694$, $\beta_3 = 7.855$, and so on, $E$ and $\rho$ are the Young's modulus and mass density of the beam material, and $V_{elas} = (E/\rho)^{1/2}$ is the elastic longitudinal wave speed that is about 21 km/s for CNTs and 5.6 km/s for $SiO_2$ nano bar. It is easily seen that the frequencies of both the vdw- and elast-GHz nanodevices fall within the GHz range as the lengths are within 100nm. In contrast to the nanobeam which has fixed and separated resonant frequencies, the frequencies of vdw-GHz oscillators are continuously tunable by changing the cross-end distance $\Delta$.

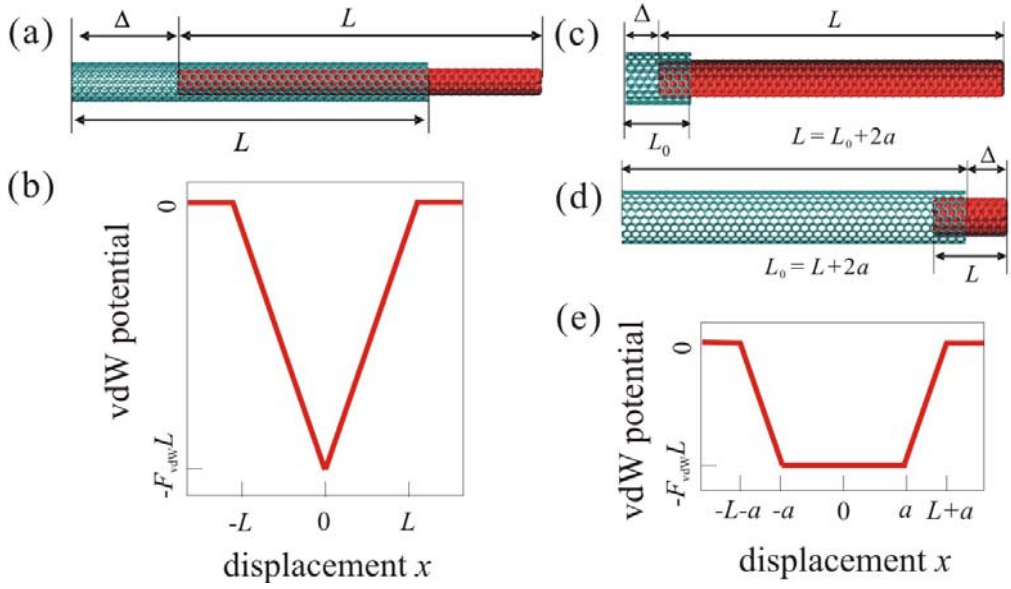

Figure 1: Possible types of DWNT GHz oscillators and corresponding intershell potentials: (a)-(b) Type-I oscillators that have equal inner and outer shell lengths and V-shaped van der Waals interaction potentials, (c)-(e) Type-II oscillators that have different inner and outer shell lengths and U-shaped potentials (schematically for system c).

Another important feature of vdw-GHz oscillators is that they permit large oscillating amplitudes; while the maximum lateral displacement, $\delta_{max}$, at the free end of a cantilever resonator is limited by the critical bending strain of the beam, $\varepsilon_{cr}$, which is often in the sub-nanometer range. It can be proved (see Supplementary Information) that $f_{elas}$ and $\delta_{max}$ are restricted by the following relation

$$f_{elas}\delta_{max} \leq \frac{\varepsilon_{cr}V_{elas}}{4\pi} \qquad (2)$$

for all resonant modes, much akin to Heisenberg's "uncertainty principle". For example, with the critical bending strain value $\varepsilon_{cr} = 0.4\%$ for MWNT beams limited by buckling instability[22] and with the structural sizes $D = 5$ nm and $L = 50$ nm, we have $f_{elas} = 5.9$ GHz and $\delta_{max} = 1.1$ nm for the fundamental resonance, and $f_{elas} = 34$ GHz and $\delta_{max} = 0.18$ nm for the second-mode resonance. Apart from the inability of direct detection using current techniques[23], these small vibration amplitudes limit its usage. The self-retraction force of a vdw-GHz oscillator is equal to[5,21] $F_{vdw} = \Pi\pi D/A_c$, where $A_c$ denotes the inner core surface area shared by a single carbon atom one, and in comparison the maximum permitted lateral force at the free end of the cantilever beam is $F_{elas} = (\pi E\varepsilon_{cr}/32)\times(D^3/L)$. For MWNTs, the ratio $F_{elas}/F_{vdw}$ is equal to 1 as $D = 19$ nm and reduces rapidly with size shrinkage. Thus, typical vdw-GHz oscillators can generate one to several orders higher work and power per volume than elas-GHz resonators.

Damping is a major obstacle towards miniaturization. An added advantage of the vdw-GHz oscillators over beam resonators lies with the much lowered degree of air damping due to their respective longitudinal and transverse motion behavior in air. Air damping was found[24] to contribute significantly to the lowering of vibration frequency and quality factor when the device size shrinks. Our analysis based on individual collisions of gas molecules in the kinetic theory of gases[25] shows that the quality factors are

$$Q_{vdw} = \frac{\pi\rho_L V_{vdw}}{\sqrt{8}K}\frac{\sqrt{\Delta L}}{(a+\frac{2}{3}\Delta)D^2}, \quad Q_{elas} = \frac{\beta^2\rho_L V_{elas}}{4K}\frac{1}{L^2}. \quad (3)$$

where $\rho_L$ denotes the mass density of the inner core or beam per unit length and $K = p(2\pi M/RT)^{1/2}$ with $M$ the molar mass of gas, $R$ the gas constant, $T$ the absolute temperature, and $p$ the gas pressure. When a vdw oscillator with a $L$-length DWNT (6,6)@(11,11) inner core and a cantilever beam resonator made from the same $L$-length DWNT (6,6)@(11,11) operate in ambient conditions where $T = 300$ K, $p = 101.325 \times 10^3$ Pa, $M = 0.028964$ kg/m$^3$, our calculations show that $Q_{vdw}$ ($10^4$~$10^5$) for Type I oscillators is two orders of magnitude higher than $Q_{elas}$ ($10^1$~$10^2$ for the basic mode) for systems of lengths greater than 50nm, as detailed in Figure 2(c). Type II oscillators have slightly lowered but similar $Q_{vdw}$. The results of the air damping $Q$-factors for beam resonators also agree with previous estimates[26]. In contrast, the thermal $Q$-factors of Type-I vdw-GHz oscillators as reviewed before are $10^1$~$10^2$ in order and those of beam GHz resonators were experimentally observed and theoretically confirmed to be 500 ~

$1000^{27, 28}$.

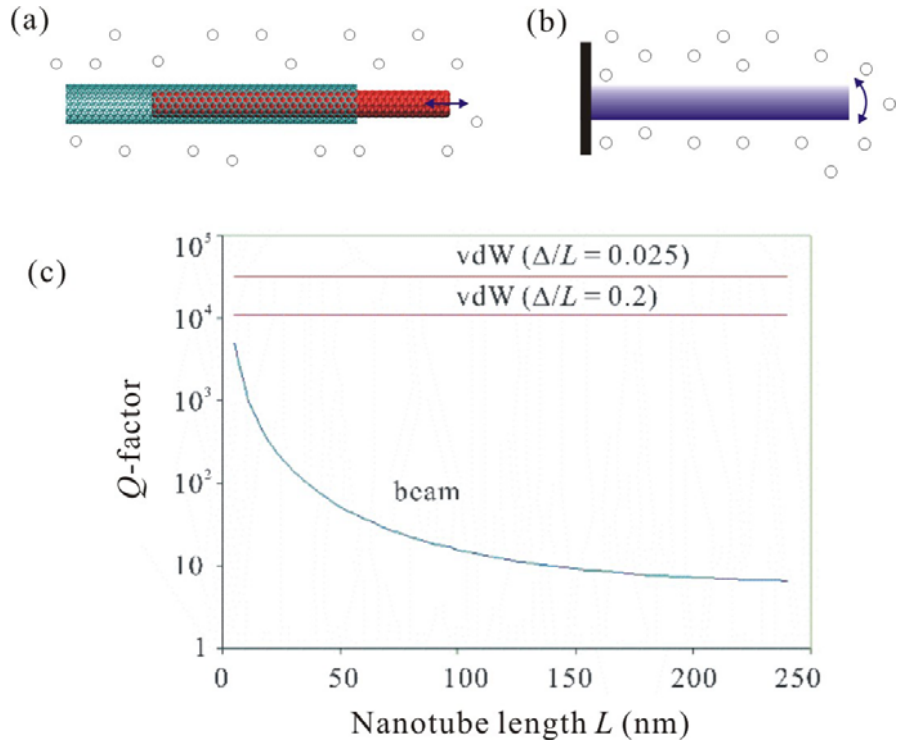

Figure 2: Quality factors under air damping: (a) vdw-GHz oscillators, (b) cantilever beam resonators, (c) the $Q$-factors of Type-I vdw-GHz oscillators are constant for given ratio $\Delta/L$ and is much higher than those of cantilever beam resonantors.

We summarize the above comparative performance features in Table 1. It is seen that Type-I vdw-GHz oscillators perform much better than nanobeam GHz resonantors in almost all the important features, except for thermal $Q$-factors.

Table 1: A comparison between externally driven van der Waals GHz oscillators and elastic nanobeam GHz resonators

| Nano-devices | Frequency | Amplitude | Work or Power | Thermal $Q$-Factor | Air damping $Q$-Factor |
|---|---|---|---|---|---|
| Beam-GHz | Quantum | Very small | Very small | Fairly high | Low |
| resonators | | (<1nm) | | (500~900) | (50~500) |
| Type-I vdw-GHz oscillators | Continuously tunable | Large | Large | Low (10~100) | High ($10^4$~$10^5$) |
| Type-II vdw-GHz oscillator | Continuously tunable | Large | Large | High (~$10^5$) | High ($10^3$~$10^4$) |

As commented previously, the way to suppress thermal damping is to reduce the cross-end effect and avoid the trans-phonon effect. It is seen from Eq. (1) or Figure 3(a) that the frequency $f_{vdw}^{II}$ of Type-II vdw-GHz oscillators as a function of the cross-end distance $\Delta$ has the vertical tangent at $\Delta = 0$ and thus may rapidly approach the GHz range with a very small $\Delta$. In addition, it is seen from Figure 3(b) that the intershell van der Waals interaction potential inside the U-shape potential well for a Type-II oscillator has much smaller fluctuation (three orders lower) than the well. Thus, if the oscillation is essentially localized inside the U-shape potential well, very low friction force and dissipation can be expected. To verify this, we carried out molecular dynamics simulation on DWNT (6,6)@(11,11) with the outer shell length $L_o$ = 2.5 nm and inner core length $L$ = 10 nm respectively. Simulation of the Dreiding force-field[29] was implemented to describe the intra- and inter-tube interaction. After structural optimization, various axial maximum speeds $v_{max}$ were imposed on the outer shell to initialize the axial oscillation. The evolution of axial speeds, taken as the signature of energy dissipation from oscillation to the phonon bath, is collectively plotted in Figure 3(c). As can be concluded from the simulation results shown in Figure 3(c), the system used avoids tube end crossing events and thus significantly reduces the energy dissipation. Specifically, when $v_{max}$ is less than 500 m/s, quality factors in the order of $10^5$ can be realized. In contrast, Type I vdw-GHz oscillators are unavoidably prone to long cross-end events, and thus have quality factors in the order of 10 to 100[12, 13, 15]. We note that the lowest phonon barrier speed is 560 m/s and the second lowest is 960 m/s[19, 19]. As $v_{max}$ is set to be 1000m/s and thus falls within the second phonon barrier zone, a drastic thermal damping is observed due to the trans-phonon effect[19, 20]. Finally, we note that a Type-II vdw-GHz oscillator has bounded frequencies. Theoretically, the frequency could reach its maximum when $f_{vdw}^{II} = V_{vdw}/(32aL)^{1/2}$ at $\Delta = a/2$, whenever $a < 2L/5$.

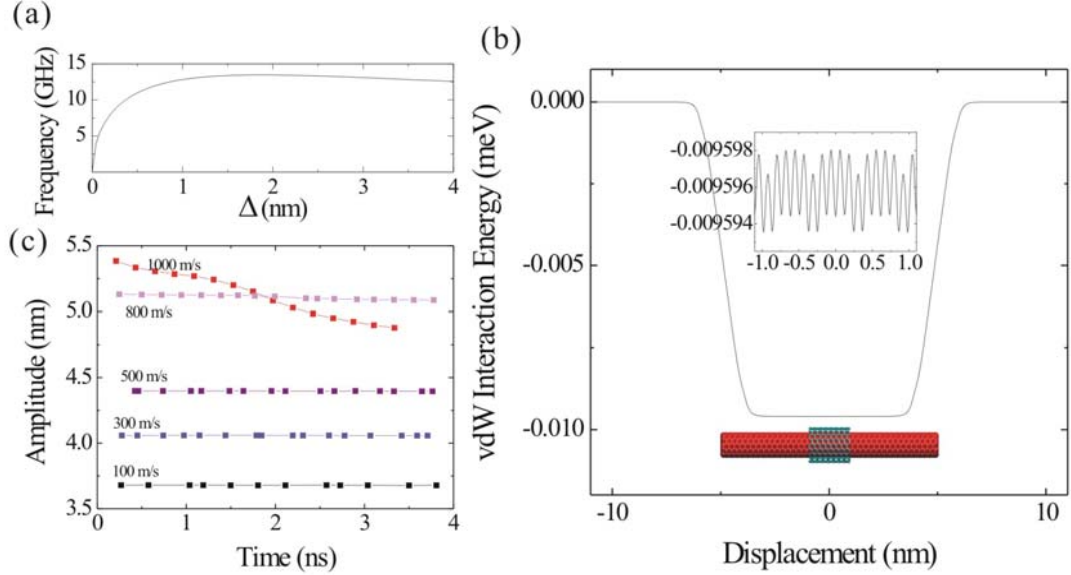

Figure 3: Frequencies and phonon dissipations of a Type II DWNT (6,6)@(11,11) oscillator with $L = 2.5$ nm, $a = 3.75$ nm, and $D = 1.49$ nm. (a) The dependence of oscillation frequencies $f_{vdw}^{II}$ on cross-end distance $\Delta$, which is limited by $L$; (b) The U-shaped potential well. It is seen that the energy fluctuation on the well bottom is three orders in magnitude lower than the well; (c) Decay of axial oscillations due to phonon dissipation. The maximum relative displacement between inner and outer shell with time is plotted for different initial axial speeds.

To tackle the last question, we conceive a NEMS based on Type-II carbon tube configuration, as illustrated in Figure 4(a). The shorter outer shell is fixed to a static platform with an electrode normal to the carbon nanotube axis. Previous experiments[10, 30] have verified that an electric field strength on the order of $10^8$ V/m can generate enough force to extrude an inner core from a one-end opened MWNT, and the force will increase by amplifying the electric field strength even though field emissions would accompany it[31]. The challenge is in verifying the oscillating phenomenon in the GHz range. Our calculation shows that applying an AC electric potential $V(t)$ of any prescribed frequency $\Omega$ not exceeding the maximum permitted value $f_{\text{vdw}}^{\text{II}}$, will enable the inner core to rapidly respond to a steady oscillating frequency at exactly the same value of $\Omega$. Our modeling calculation is carried out for a Type-II DWNT (6,6)@(11,11) oscillator with inner tube length $L$ = 2.5 nm and outer tube length $L_o$ = 10 nm (or $a$ = 3.75 nm), as studied previously. The van der Waals force is taken as the analytical solution provided[5,21], and as such varies in a piecewise manner with location. Damping is taken to be that due to collision of gas molecules and is dependant on velocity. Finally a time varying sinusoidal excitation axial force, $A\sin(2\pi\Omega t)$, is applied to simulate the loading by an external field[10, 30]. These three forces are equated with the inertia of the core, and the resulting expression solved in a finite difference algorithm in the time domain. The dynamical response of the displacement peak $x_{\max}$ to altering the frequency of the driving force is demonstrated in Figure 4(b). In the simulation, a sinusoidal force of amplitude $A$ = 10 pN was applied, as the frequency $\Omega$ was varied every 1.2 μs so that it took sequential values of 4.0 GHz, 4.4 GHz, 4.8 GHz and 4.0 GHz. A time step of 40 fs was used for the analysis with the initial displacement as 4 nm. The plot in Figure 4(b) clearly indicates amplitude change in response to the altering frequency. We note that the external force amplitude $A$ = 10 pN is much lower than the previously estimated van der Waals retraction force $F_{\text{vdw}} = \Pi\pi D/A_c \approx$ 1 nN. More importantly, our calculations show that the value of $A$ and damping do not affect the response oscillation frequency $f_{\text{vdw}}^{\text{II}}$, which is exactly equal to the applied $\Omega$, although they have influence on the oscillation amplitude. The above properties, revealed based on our simulations may play a key role in experimentally realizing vdw-GHz oscillators. For example, SEM observation of a fuzzy image of the oscillator with an elongated length, approximately $2L - L_0$, with the frequency exactly equal to the applied $\Omega$, should be an important guide when conducting experimental verification of GHz oscillation.

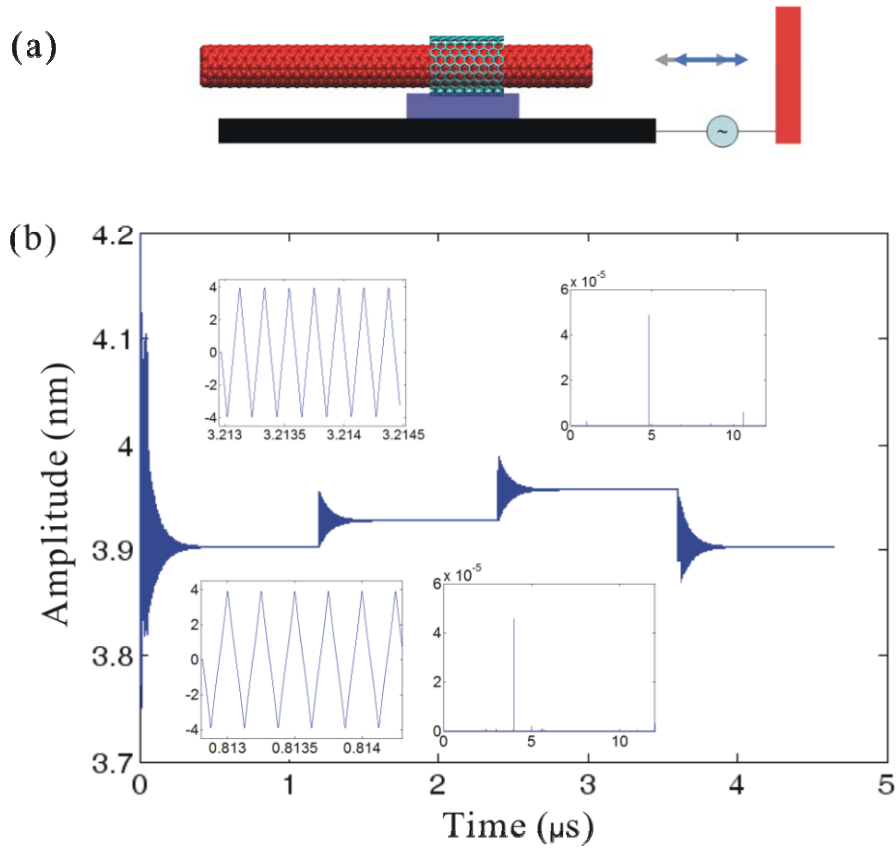

Figure 4: Continuous frequency controllable NEMS based on a Type II DWNT oscillator. (a) Illustrative setting of vdw-GHz oscillator driven by a time varying sinusoidal excitation axial force. (b) Plot of the maximum driven displacements against time for sequential driving frequencies of 4.0 GHz, 4.4 GHz, 4.8 GHz and 4.0 GHz under forcing amplitudes of 10 pN. The left Inserts show the moving distance time traces as $\Omega$ = 4.8 GHz (above) and 4.0 GHz (below), and the right Inserts show the corresponding fast Fourier transformations.

In summary, the Type II GHz oscillators based on multiwalled carbon nanotubes and van der Waals interaction proposed in this Letter have significant advantages over nanobeam based GHz resonators in most important features, such as power density, work density, displacement amplitude, thermal damping *Q*-factor, air damping *Q*-factor, and continuous frequency tunability. We hope these merits will help to realize such NEMS sooner and prompt their study for various applications. With the discovery of self-retracting motion in graphite[32], we envisage application of the same principles outlined in this Letter to conceive and realize NEMS with similar advantages based on graphite as well as graphite-carbon nanotubes.

**Acknowledgements** This work was supported to ZQS by the National Science Foundation of China (NSFC) through Grants 10121202, 10332020 and 50518003, and National Basic Research Program of China through Grant 2007CB936803.

**Reference**

[1] M. C. Roco, Journal of Nanoparticle Research **7**, 707 (2005).

[2] K. E. Drexler, *Nanosystems: molecular machinery, manifacturing & computation* (Wiley, New York, 1992).

[3] M. F. Yu, O. Lourie, M. J. Dyer, K. Moloni, T. F. Kelly and R. S. Ruoff, Science 287, 637 (2000); M. F. Yu, B. I. Yakobson, and R. S. Ruoff, Journal of Physical Chemistry B **104**, 8764 (2000).

[4] J. Cumings and A. Zettl, Science **289**, 602 (2000).

[5] Q. S. Zheng and Q. Jiang, Physical Review Letters **88**, 045503 (2002);.

[6] J. R. Minkel, Physical Review Focus (2002).

[7] X. M. H. Huang, C. A. Zorman, M. Mehregany, *et al.*, Nature **421**, 496 (2003).

[8] K. Jensen, C. Girit, W. Mickelson, *et al.*, Physical Review Letters **96**, 215503 (2006).

[9] L. X. Dong, B. J. Nelson, T. Fukuda, *et al.*, IEEE Transactions on Automation Science and Engineering **3**, 228 (2006).

[10] A. Subramanian, L. X. Dong, J. Tharian, *et al.*, Nanotechnology **18**, 075703 (2007).

[11] P. Mohanty, D. A. Harrington, K. L. Ekinci, *et al.*, Physical Review B **66**, 085416 (2002).

[12] W. L. Guo, Y. F. Guo, H. Gao, Q.S. Zheng, *et al.*, Physical Review Letters **91**,

125501 (2003).

[13] Y. Zhao, C. C. Ma, G. H. Chen, *et al.*, Physical Review Letters **91**, 175504 (2003).

[14] J. L. Rivera, C. McCabe, and P. T. Cummings, Nano Letters **3**, 1001 (2003).

[15] P. Tangney, S. G. Louie, and M. L. Cohen, Physical Review Letters **93**, 065503 (2004).

[16] F. F. Abraham, H. J. Gao, and Q. S. Zheng, unpublished (2002).

[17] W. L. Guo, W. Y. Zhong, Y. T. Dai, *et al.*, Physical Review B **72**, 075409 (2005).

[18] L. H. Wong, Y. Zhao, G. H. Chen, *et al.*, Applied Physics Letters **88**, 183107 (2006).

[19] Q. S. Zheng, in *3rd International Conference on Materials for Advanced Technologies*, Singapore, (2005); Z. Xu, Q.S. Zheng, Q. Jiang, *et al.*, http://arxiv.org/abs/cond-mat/0709.0989 (2007).

[20] P. Tangney, M. L. Cohen, and S. G. Louie, Physical Review Letters **97**, 195901 (2006).

[21] Q. S. Zheng, J. Z. Liu, and Q. Jiang, Physical Review B **65**, 245409 (2002)

[22] J. Z. Liu, Q. Zheng, and Q. Jiang, Physical Review Letters **86**, 4843 (2001); J. Z. Liu, Q. Zheng, and Q. Jiang, Physical Review B **67**, 075414 (2003).

[23] M. D. LaHaye, O. Buu, B. Camarota, *et al.*, Science **304**, 74 (2004).

[24] C. L. Zhang, G. S. Xu, and Q. Jiang, Math. Mech. Solids **8**, 315 (2003).

[25] A. Neild, T.W. Ng, Z.P. Xu, Q. S. Zheng, (to be submitted, see also Supplementary Information).

[26] M. Michael James and H. H. Brian, Applied Physics Letters **91**, 103116 (2007).

[27] P. Poncharal, Z. L. Wang, D. Ugarte, *et al.*, Science **283**, 1513 (1999).

[28] H. Jiang, M. F. Yu, B. Liu, *et al.*, Physical Review Letters **93**, 185501 (2004).

[29] Y. Guo, N. Karasawa, and W. A. Goddard, Nature **351**, 464 (1991).

[30] V. V. Deshpande, H. Y. Chiu, H. W. C. Postma, *et al.*, Nano Lett. **6**, 1092 (2006).

[31] W. Wei, J. Kaili, W. Yang, *et al.*, Nanotechnology, **17**, 1994 (2006).

[32] Q.S.. Zheng, B. Jiang, S.P. Liu, *et al.*, Self-retracting motion of graphite microflakes. Physical Review Letters (under review).